\documentclass[conference,letterpaper]{IEEEtran}
\usepackage[utf8]{inputenc} 
\usepackage[T1]{fontenc}
\usepackage{url}
\usepackage{cite}
\usepackage[cmex10]{amsmath} 
\usepackage{amssymb,amsfonts}
\usepackage{bbm}
\usepackage{algorithm,algorithmic}
\usepackage{graphicx}
\usepackage{textcomp}
\usepackage{xcolor}
\usepackage{float}
\usepackage{caption}
\usepackage{cases}
\usepackage{stfloats}
\newtheorem{theorem}{\rm{\textbf{Theorem}}}
\newtheorem{lemma}{\rm{\textbf{Lemma}}}
\newtheorem{definition}{\rm{\textbf{Definition}}}
\newtheorem{proposition}{\rm{\textbf{Proposition}}}
\newtheorem{corollary}{\rm{\textbf{Corollary}}}

\newtheorem{remarkex}{\rm{\textbf{Remark}}}
\graphicspath{{figures/}}

\begin{document}
\title{Strong Privacy and Utility Guarantee: Over-the-Air Statistical Estimation} 

\author{%
  \IEEEauthorblockN{Wenhao~Zhan}
  \IEEEauthorblockA{Tsinghua Universiry\\
  	                Dept. of Electronic Engineering\\
                    100084 Beijing, China\\
                    Email: zhanwh17@mails.tsinghua.edu.cn}
}

\maketitle

\begin{abstract}
We consider the privacy problem of statistical estimation from distributed data, where users communicate to a central processor over a Gaussian multiple-access channel(MAC). To avoid the inevitable sacrifice of data utility for privacy in digital transmission schemes, we devise an over-the-air estimation strategy which utilizes the additive nature of MAC channel. Using the mutual information between the channel outputs and users' data as the metric, we obtain the privacy bounds for our scheme and validate that it can guarantee strong privacy without incurring larger estimation error. Further, to increase the robustness of our methods, we adjust our primary schemes by adding Gaussian noises locally and derive the corresponding minimax mean squared error under conditional mutual information constraints. Comparing the performance of our methods to the digital ones, we show that the minimax error decreases by $O(\frac{1}{n})$ in general, which suggests the advantages of over-the-air estimation for preserving data privacy and utility.        
\end{abstract}


\section{Introduction}
Nowadays the prosperity of mobile devices and rapid development of Internet of things have created massive amount of data in distributed edge devices, which can play an important role in applications such as medical care\cite{xu2019federated}, purchase recommendation\cite{hegedHus2019decentralized} and keyboard prediction\cite{hard2018federated}. Typically these data need to be transmitted to a central processor for inference and learning. However, due to the sensitive information contained in them, the  data collected by companies will cause privacy leakage if not processed properly and prohibit users from sharing. Therefore, how to solve statistical inference problems from distributed data while guaranteeing users' privacy has become a new challenge for researchers.

Statistical estimation is one of the most common tasks in distributed inference setting. There have been extensive research studying the performance of distributed estimation problems under privacy constraints \cite{duchi2013local,sheffet2018locally,acharya2018distributed,barnes2020fisher,erlingsson2014rappor,kairouz2016discrete,ye2018optimal}. They usually only consider local differential privacy (LDP) \cite{duchi2013local}, a private mechanism that has been applied by several major technology organizations including Apple, Google and Microsoft\cite{cormode2018privacy}. However, as the information-theoretic view for privacy and privacy funnel problems \cite{makhdoumi2014information} become popular recently, mutual information has also been considered as a promising privacy metric in many works \cite{calmon2015fundamental,asoodeh2015maximal,liao2017hypothesis,sreekumar2019optimal} because LDP is often too strict to implement in practice. \cite{wang2016relation,cuff2016differential} investigated the relationships between LDP and mutual information privacy and showed that mutual information constraints are in fact sandwiched between pure LDP and approximate LDP. This further validated the rationality of mutual information privacy. 

Another limitation of existing literature is that they usually assume a simplified digital communication scheme where each user sends the outputs of private mechanisms independently and the processor receives their data one by one without error and interference. Lately over-the-air machine learning was proposed and has attracted much attention. Over-the-air means that each user simply scales and transmits their data to the MAC channel simultaneously in an analog, uncoded fashion. By leveraging the additive nature of MAC channel, analog transmission schemes have shown significant performance enhancement and stronger privacy guarantee compared to traditional digital communication schemes in federated learning problems \cite{amiri2020machine,seif2020wireless}. Nevertheless, no work has been done on the performance of over-the-air transmission for statistical estimation under privacy constraints, which should be an interesting topic considering the advantages of analog transmission schemes in machine learning.

In this work we consider the distributed estimation problems under mutual information privacy. We study the minimax mean squared estimation error for three common models: Gaussian location model, product Bernoulli model and sparse Bernoulli model. To satisfy mutual information constraints, we devise over-the-air estimation schemes and validate that they can achieve arbitrary privacy requirements without sacrificing the data utility. We also investigate the conditional mutual information constraints. By adding local Gaussian noises, our proposed schemes can be adjusted to fit in more general models and meet the demands of this more robust privacy metric, displaying much lower estimation error compared to digital schemes.

The rest of the paper is organized as follows: we formulate the over-the-air statistical estimation problem in Section II. In Section III we propose the minimax optimal estimation schemes and derive their mutual information privacy guarantee and minimax estimation error. In Section IV we consider the conditional mutual information constraints and adjust out primary schemes to be more robust. Then we validate the enhanced performance of our  revised schemes over traditional digital schemes under conditional mutual information privacy. The conclusion is drawn in Section V.

\section{Problem Formulation}
We consider a wireless communication system with $n$ users and a central processor. Each user has an i.i.d sample $U_i$ which follows an unknown distribution $p_{\theta}$ belonging to a parameterized family of distributions $\mathcal{P}=\{p_{\theta}\vert\theta\in\Theta\subseteq\mathbb{R}^d\}$. To protect the privacy, user $i$ first processes his or her data with encoding function $f(\cdot)$ and obtains the output $X_i\triangleq(X_{i1},X_{i2},\cdots,X_{is})=f(U_i)$ which satisfies the privacy constraints. Then the users share a Gaussian MAC channel to transmit their symbols to the processor. In channel use $j$($1\leq j\leq s$), every user $i$ sends the symbol $X_{ij}$ to the channel simultaneously for all $1\leq i\leq n$, and the processor will receive $Y_j$, the summation of all signals with channel noises,
\begin{equation}
Y_j = \sum_{i=1}^nX_{ij}+Z_j,
\end{equation}
where $Z_j\sim\mathcal{N}(0,\sigma_0^2)$. To model the limited battery capacity in edge devices, we assume a power constraint on each user:
\begin{equation}
\frac{1}{s}\sum_{j=1}^{s}\mathbb{E}_{\theta,f}[X_{ij}^2]\leq P,\text{ for all } 1\leq i\leq n,
\end{equation}
where $\mathbb{E}_{\theta,f}[\cdot]$ denotes the expectation with respect to $p_{\theta}$ and $f(\cdot)$.

After $s$ channel uses, the processor will attempt to estimate the unknown parameter $\theta$ from $Y\triangleq(Y_1,\cdots,Y_s)$. Suppose the estimator is $\hat{\theta}(Y)$, then let $R_{\theta}(\hat{\theta},f)$ denote the mean squared estimation error for encoding function $f$ and estimator $\hat{\theta}$ under $p_{\theta}$,
\begin{equation}
R_{\theta}(\hat{\theta},f)=\mathbb{E}_{\theta,f}[\Vert\hat{\theta}-\theta\Vert_2^2].
\end{equation}
Our goal is to minimize the minimax estimation error $R(\hat{\theta},f)=\sup_{\theta\in\Theta}R_{\theta}(\hat{\theta},f)$ and suppose the minimax optimal pair of encoding function and estimator is $(\hat{\theta}^*,f^*)$.

In this work we mainly focus on the following three distribution families. They are all common models and widely applied in other areas such as machine learning.
\begin{definition}
Gaussian location model refers to the distribution family $p_{\theta}=\mathcal{N}(\theta,\sigma^2I_d)$, where $\Theta=\{\theta\in\mathbb{R}^d:\Vert\theta\Vert_2\leq B\sqrt{d}\}$ for some known $\sigma,B>0$.
\end{definition}

\begin{definition}
	Product Bernoulli model refers to the distribution family $p_{\theta}=\prod_{k=1}^{d}\text{Bernoulli}(\theta_k)$, where $\theta\triangleq(\theta_1,\cdots,\theta_d)$ and $\Theta=[0,1]^d$.
\end{definition}

\begin{definition}
	$m$-Sparse Bernoulli model refers to the distribution family $p_{\theta}=\prod_{k=1}^{d}\text{Bernoulli}(\theta_k)$, where $\theta\triangleq(\theta_1,\cdots,\theta_d)$ and $\Theta=\{\theta\in[0,1]^d\vert\sum_{k=1}^d\theta_k\leq m\}$.
\end{definition}
\section{Over-the-Air Estimation: Free Mutual Information Privacy}
In this section we first propose the minimax optimal estimation schemes (i.e., the optimal pairs of encoding function and estimator) for the above three models without considering the privacy constraints. Then we will investigate the mutual information bounds between the channel output $Y$ and each data $U_i$ that these schemes satisfy, denoted by $I_{\theta}(Y;U_i)$.
\subsection{Minimax Optimal Over-the-Air Estimation}
First we want to obtain the minimax optimal estimation schemes via over-the-air transmission without the privacy constraints. Since we do not have to protect privacy, we can only consider deterministic and invertible encoding function. \cite{} has studied this problem for Gaussian location model and product Bernoulli model, so we introduce them here as Lemma~\ref{Gaussian minimax} and Lemma~\ref{Bernoulli minimax}. Notice that in our work we focus on the scenario where channel uses $s=d$. When $s>d$ we can use repetition to obtain better performance.
\begin{lemma}
\label{Gaussian minimax}
For Gaussian location model, suppose each user uses the same linear encoding function $f(u)=\alpha u+\beta$ for some $\alpha\in\mathbb{R},\beta\in\mathbb{R}^d$ and the processor uses arbitrary estimator, then the minimax optimal over-the-air scheme is:
\begin{equation}
f^*(u)=\sqrt{\frac{P}{B^2+\sigma^2}}u, \hat{\theta}^*(Y)=\frac{1}{n}\sqrt{\frac{B^2+\sigma^2}{P}}Y,
\end{equation}
and the corresponding minimax estimation error is:
\begin{equation}
R(\hat{\theta}^*,f^*)=\frac{d\sigma^2}{n}\left[1+\frac{\sigma_0^2}{nP}\left(1+\frac{B^2}{\sigma^2}\right)\right].
\end{equation}
\end{lemma}

\begin{lemma}
\label{Bernoulli minimax}
For product Bernoulli model, suppose each user uses arbitrary encoding functions and the processor uses affine estimators, then the minimax optimal over-the-air encoding function is:
\begin{equation}
f^*(u)=\sqrt{P}(2u-1), 
\end{equation}
and the estimator is $\hat{\theta}^*(Y)=\alpha^*Y+\beta^*\mathbf{1}$, where $\beta^*=\frac{1}{2}$ and
\begin{equation}
\alpha^*=
\begin{cases}
\frac{1}{2\sqrt{nP}(\sqrt{n}+1)}, &\text{ if } \sigma_0^2\leq n^{\frac{3}{2}}P,\\
\frac{n\sqrt{P}}{2(\sigma_0^2+n^2P)}, &\text{ if } \sigma_0^2\geq n^{\frac{3}{2}}P.
\end{cases}
\end{equation}
and the corresponding minimax estimation error is:
\begin{equation}
R(\hat{\theta}^*,f^*)=
\begin{cases}
\frac{d}{4(\sqrt{n}+1)^2}(1+\frac{\sigma_0^2}{nP}), & if \sigma_0^2\leq n^{\frac{3}{2}}P,\\
\frac{d}{4}\cdot\frac{1}{1+n\cdot\frac{nP}{\sigma_0^2}}, & if \sigma_0^2\geq n^{\frac{3}{2}}P.
\end{cases}
\end{equation}
\end{lemma}

The sparsity of a model can usually help us reduce estimation error and communication cost. The following Theorem~\ref{sparse minimax} shows that for sparse product Bernoulli model we can obtain a much smaller minimax error than ordinary product Bernoulli model, which scales with the sparsity level $m$ instead of $d$ when the model is sufficiently sparse.
\begin{theorem}
	\label{sparse minimax}
	For $m$-sparse product Bernoulli model, suppose each user uses arbitrary encoding functions and the processor uses affine estimators. Then,\\
	(1) when $\frac{m}{d}\geq\frac{1}{2}$: the minimax optimal over-the-air estimation scheme and corresponding estimation error are the same as product Bernoulli model. \\
	(2) when $\frac{m}{d}\leq\frac{1}{2}$: the minimax optimal over-the-air estimation function is 
	\begin{equation}
	[f^*(u)]_j=
	\begin{cases}
	-\sqrt{\frac{m}{d-m}P}, & if [u]_j=0,\\
	\sqrt{\frac{d-m}{m}P}, & if [u]_j=1.
	\end{cases}
	\end{equation}
	 where $[\cdot]_j$ is the $j$-th element of the vector, and the estimator is $\hat{\theta}^*(Y)=\alpha^*Y+\beta^*\mathbf{1}$ where
	\begin{equation}
	\begin{split}
	 \beta^*=(1-\frac{2m}{d})\frac{n\sqrt{P}}{2}(\sqrt{\frac{d-m}{m}}+\sqrt{\frac{m}{d-m}})\alpha^*\\
	 +\frac{m}{d}-\frac{n\sqrt{P}(\sqrt{\frac{d-m}{m}}-\sqrt{\frac{m}{d-m}})}{2}\alpha^*. 
	 \end{split}
	 \end{equation}
	 If $\frac{d^2}{4m(d-m)}\sigma_0^2\leq n^{\frac{3}{2}}P$, then
	\begin{equation}
	\alpha^*=\frac{1}{\sqrt{nP}(\sqrt{\frac{d-m}{m}}+\sqrt{\frac{m}{d-m}})(\sqrt{n}+1)},
    \end{equation}
    and the minimax estimation error is
    \begin{align}
    &R(\hat{\theta}^*,f^*)\notag\\
    =&\frac{m}{(\sqrt{n}+1)^2}\cdot(\frac{d-m}{d}
    +\frac{d\sigma_0^2}{mnP(\sqrt{\frac{d-m}{m}}+\sqrt{\frac{m}{d-m}})^2}).
    \end{align}
    
	 If $\frac{d^2}{4m(d-m)}\sigma_0^2\geq n^{\frac{3}{2}}P$, then
\begin{equation}
\alpha^*=\frac{m(d-m)n\sqrt{P}(\sqrt{\frac{d-m}{m}}+\sqrt{\frac{m}{d-m}})}{d^2\sigma_0^2+m(d-m)n^2P(\sqrt{\frac{d-m}{m}}+\sqrt{\frac{m}{d-m}})^2},
\end{equation}
and the minimax estimation error is
\begin{align}
&R(\hat{\theta}^*,f^*)\notag\\
=&m\cdot\frac{1}{\frac{d}{d-m}+n\cdot\frac{mnP(\sqrt{\frac{d-m}{m}}+\sqrt{\frac{m}{d-m}})^2}{d\sigma_0^2}}.
\end{align}
\end{theorem}
\begin{IEEEproof}
Here we provide a sketch of proof. For more details please see the appendices. We first consider the minimax optimal schemes for a subset of encoding functions parameterized by $C$:
\begin{equation*}
f_C(u)=C(2u-1).
\end{equation*}
and we assume the estimator is $\hat{\theta}_{\alpha,\beta}=\alpha Y+\beta$. Then the minimax estimation error can be expressed as follows:
\begin{align*}
&\sup_{\theta}R_{\theta}(\hat{\theta}_{\alpha,\beta},f_C)=[4(n^2-n)C^2\alpha^2-4nC\alpha+1]\sum_{j=1}^{d}\theta_j^2\notag\\
&+[4nC^2\alpha^2+4nC\alpha\beta-2\beta-4n^2C^2\alpha^2+2nC\alpha]\sum_{j=1}^{d}\theta_j\notag\\
&+d[\alpha^2\sigma_0^2+(\beta-nC\alpha)^2].
\end{align*}

This is a quadratic function of $d$ variables $\theta_1,\cdots,\theta_d$ and it is difficult to obtain its maximum point directly. Therefore we first utilize the following inequality to reduce the number of variables:
\begin{equation}
\label{key inequality}
\frac{(\sum_{j=1}^d\theta_j)^2}{d}\leq\sum_{j=1}^d\theta_j^2\leq\sum_{j=1}^d\theta_j
\end{equation} 
The left side takes equality when all $\theta_i$ is equal and the right side takes equality when all $\theta_i$ takes values in $0$ or $1$. Then we can consider the following different cases:

(1)$4(n^2-n)C^2\alpha^2-4nC\alpha+1\geq0$: since the coefficient of $\sum_{j=1}^d\theta_j^2$ is larger than 0, we can utilize the right side of Inequality~(\ref{key inequality}) and the minimax error can thus be written as:
\begin{align*}
&\sup_{\theta}R_{\theta}(\hat{\theta}_{\alpha,\beta},f_C)=[(2nC\alpha-1)(2\beta-1)]\sum_{j=1}^d\theta_j\\
&+d[\alpha^2\sigma_0^2+(\beta-nC\alpha)^2].
\end{align*}
Now this function is only a linear function of $\sum_{j=1}^d\theta_j$ and we can handle it more easily. The following lemma characterizes the optimal scheme in this scenario:
\begin{lemma}
\label{prooflemma1}
	When $4(n^2-n)C^2\alpha^2-4nC\alpha+1\geq0$, the minimax optimal scheme for $\alpha,\beta$ is
	\begin{equation}
	\alpha^*=
	\begin{cases}
	\min\{\frac{1}{2\sqrt{n}C(\sqrt{n}+1)},\frac{nC}{2(\sigma_0^2+n^2C^2)}\},& if \frac{m}{d}\geq\frac{1}{2},\\
	\min\{\frac{1}{2\sqrt{n}C(\sqrt{n}+1)},\frac{2m(d-m)nC}{d^2\sigma_0^2+4m(d-m)n^2C^2)}\},& if \frac{m}{d}\leq\frac{1}{2}.
	\end{cases}
	\end{equation}
	\begin{equation}
	\beta^*=
	\begin{cases}
	\frac{1}{2},& if \frac{m}{d}\geq\frac{1}{2},\\
	(1-\frac{2m}{d})nC\alpha^*+\frac{m}{d},& if \frac{m}{d}\leq\frac{1}{2}.
	\end{cases}
	\end{equation}
\end{lemma}

(2)$4(n^2-n)C^2\alpha^2-4nC\alpha+1\leq0$: Similarly, we can utilize the left side of Inequality~(\ref{key inequality}) to obtain the simplified minimax error in this case:
\begin{align*}
&\sup_{\theta}R_{\theta}(\hat{\theta}_{\alpha,\beta},f_C)=[4(n^2-n)C^2\alpha^2-4nC\alpha+1]\frac{(\sum_{j=1}^{d}\theta_j)^2}{d}\notag\\
&+[4nC^2\alpha^2+4nC\alpha\beta-2\beta-4n^2C^2\alpha^2+2nC\alpha]\sum_{j=1}^{d}\theta_j\notag\\
&+d[\alpha^2\sigma_0^2+(\beta-nC\alpha)^2].
\end{align*}
This function is a quadratic function of $\sum_{j=1}^d\theta_j$. The following lemma displays the optimal scheme in this case.
\begin{lemma}
\label{prooflemma2}
	When $4(n^2-n)C^2\alpha^2-4nC\alpha+1\leq0$, the minimax optimal scheme for $\alpha,\beta$ is
	\begin{equation}
	\alpha^*=\frac{1}{2\sqrt{n}C(\sqrt{n}+1)},
	\end{equation}
	\begin{equation}
	\beta^*=
	\begin{cases}
	\frac{1}{2},& if \frac{m}{d}\geq\frac{1}{2},\\
	(1-\frac{2m}{d})nC\alpha^*+\frac{m}{d},& if \frac{m}{d}\leq\frac{1}{2}
	\end{cases}
	\end{equation}
\end{lemma}

From Lemma~\ref{prooflemma1} and \ref{prooflemma2}, we can see that the optimal scheme in the second case is included in the first case. Thus the scheme proposed in Lemma~\ref{prooflemma1} is exactly the optimal minimax estimation scheme under $f_C(\cdot)$.  Finally we need to convert arbitrary deterministic encoding functions into $f_C$, which is shown in the following lemma:
\begin{lemma}
\label{prooflemma3}
	Suppose the encoding fucntion is as follows:
	\begin{equation}
	[f(u)]_j=
	\begin{cases}
	A &\text{ if }[u]_j=0,\\
	B &\text{ if }[u]_j=1.
	\end{cases}
	\end{equation}
	Then $\sup_{\theta}R(\theta;f,\hat{\theta}_{\alpha,\beta})=\sup_{\theta}R(\theta;f_C,\hat{\theta}')$, where $C=\frac{B-A}{2}$ and $\hat{\theta}'(Y)=\hat{\theta}(Y+n\frac{A+B}{2})$.
\end{lemma}

Lemma~\ref{prooflemma3} tells us that for each pair $(\hat{\theta},f)$, we can find an equivalent pair $(\hat{\theta}',f_C)$. Since $R(\hat{\theta}_{\alpha^*,\beta^*},f_C)$ decreases with $\vert C\vert$, we want the largest equivalent $\vert C^*\vert$ that satisfies the power constraint to attain optimal performance, which is considered in the following lemma.
\begin{lemma}
\label{prooflemma4}
	For the optimal encoding function, when $\frac{m}{d}\geq\frac{1}{2}$, we have
	\begin{equation} 
	A^*=-\sqrt{P},B^*=\sqrt{P},C^*=\sqrt{P}.
	\end{equation}
	
	When $\frac{m}{d}\leq\frac{1}{2}$, we have
	\begin{scriptsize}
		\begin{equation} 
		A^*=-\sqrt{\frac{m}{d-m}P},B^*=\sqrt{\frac{d-m}{m}P},C^*=\frac{1}{2}\sqrt{(\frac{d-m}{m}+\frac{m}{d-m})P}.
		\end{equation}
	\end{scriptsize}
\end{lemma}

Substituting Lemma~\ref{prooflemma4} and Lemma~\ref{prooflemma3} into Lemma~\ref{prooflemma1}, we can obtain Theorem~\ref{sparse minimax}.
\end{IEEEproof}
\subsection{Mutual Information Bounds for Over-the-Air Estimation}
Now that we have derived the minimax optimal schemes, we need to validate the privacy guarantee they provide without using additional private mechanisms. The key idea is to consider other users' signals as noises when deriving the upper bounds $I_{\theta}(Y;U_i)$ for user $i$. First for Gaussian location model we have the following proposition:
 \begin{proposition}
 	\label{mutual Gaussian}
 	For Gaussian location model, suppose each user uses the optimal schemes in Lemma~\ref{Gaussian minimax}, then the mutual information
 	\begin{equation}
 	I_{\theta}(Y;U_i)\leq\frac{d}{2}\frac{1}{n-1+\frac{\sigma_0^2(B^2+\sigma^2)}{P\sigma^2}}, \forall 1\leq i\leq n.
 	\end{equation}
 \end{proposition}  

 \begin{IEEEproof}
 	Notice each user's signal can be viewed as a combination of true parameters and Gaussian observation noises, i.e., $X_i=\sqrt{\frac{P}{B^2+\sigma^2}}\theta + Z_{user,i}$ where $Z_{user,i}\sim\mathcal{N}(0,\sigma^2I_d)$. Denote $Z\in\mathbb{R}^d$ to be the channel noise vector $(Z_1,\cdots,Z_d)$, then we have: 
 	\begin{align*}
 	Y=&X_i+(n-1)\sqrt{\frac{P}{B^2+\sigma^2}}\theta+Z+\sum_{l\neq i}Z_{user,l}\\
 	&=X_i+(n-1)\sqrt{\frac{P}{B^2+\sigma^2}}\theta+Z_{up},
 	\end{align*}
 	where $Z_{up}\sim\mathcal{N}(0,\sigma_0^2+(n-1)\frac{P}{\sigma^2+B^2}\sigma^2I_d)$. Considering that $X_i\sim\mathcal{N}(\sqrt{\frac{P}{B^2+\sigma^2}}\theta,\frac{P}{B^2+\sigma^2}\sigma^2I_d)$, we have:
 	\begin{equation}
 	I_{\theta}(Y_{j};X_{ij})= \frac{1}{2}\log\left(\frac{1}{n-1+\frac{\sigma_0^2(B^2+\sigma^2)}{P\sigma^2}}+1\right)
 	\end{equation}
 	Since $d$ channel uses are independent of each other, we have:
 	\begin{align*}
 	&I_{\theta}(Y;X_{i})\notag\\
 	\leq&\frac{d}{2}\log\left(\frac{1}{n-1+\frac{\sigma_0^2(B^2+\sigma^2)}{P\sigma^2}}+1\right)\notag\\
 	\leq&\frac{d}{2}\frac{1}{n-1+\frac{\sigma_0^2(B^2+\sigma^2)}{P\sigma^2}}.
 	\end{align*}
 	Notice that the encoding function is determined and invertible, therefore $I_{\theta}(Y;U_{i})=I_{\theta}(Y;X_{i})\leq\frac{d}{2}\frac{1}{n-1+\frac{\sigma_0^2(B^2+\sigma^2)}{P\sigma^2}}$
 \end{IEEEproof} 
 
 Similarly, for product Bernoulli model we can also bound the mutual information:
 \begin{proposition}
 	\label{mutual Bernoulli}
 	For product Bernoulli model, suppose each user uses the optimal schemes in Lemma~\ref{Bernoulli minimax}, then the mutual information
 	\begin{equation}
 	I_{\theta}(Y;U_i)\leq\frac{d}{n}.
 	\end{equation}
 \end{proposition}
 \begin{remarkex}
	Notice that for $m$-sparse product Bernoulli model we can have a similar conclusion, i.e., $I_{\theta}(Y;U_{i})\leq \frac{d}{n}$.
\end{remarkex}
 \begin{IEEEproof}
 	Denote $\tilde{Y}=\sum_{i=1}^nX_i$ to be the superposition of all signals without channel noise, then we have $I_{\theta}(Y;U_i)\leq I_{\theta}(\tilde{Y};U_i)=I_{\theta}(\tilde{Y};X_i)$.
 	By the definition of mutual information, we have: 
 	\begin{align}
 	&I_{\theta}(\tilde{Y}_j;X_{ij})\notag\\
 	=&\theta_j\sum_{y_j=1}^np(y_j\vert u_{ij}=1)\log\frac{p(y_j\vert u_{ij}=1)}{p(y_j)}\notag\\
 	&+(1-\theta_j)\sum_{y_j=0}^{n-1}p(y_j\vert u_{ij}=0)\log\frac{p(y_j\vert u_{ij}=0)}{p(y_j)}\label{Bernoulli mutual}.
 	\end{align}
 	To facilitate our proof, we introduce a lemma first:
 	\begin{lemma}
 	\label{prooflemma5}
 	Let $V$ ba any nonnegative random variable with mean $\mu$ and variance $\omega^2$, then
 	\begin{equation} 
 	\mathbb{E}[V\ln V]\leq\mu\ln\frac{\omega^2+\mu^2}{\mu}.
 	\end{equation}
 	\end{lemma}
 	Now we bound the first item in (\ref{Bernoulli mutual}):
 	\begin{align} 
 	&\theta_j\sum_{y_j=1}^np(y_j\vert u_{ij}=1)\log\frac{p(y_j\vert u_{ij}=1)}{p(y_j)}\notag\\
 	=&\theta_j\sum_{y_j=1}^n\theta_j^{y_j-1}(1-\theta_j)^{n-y_j}C_{n-1}^{y_j-1}\log\frac{\theta_j^{y_j-1}(1-\theta_j)^{n-y_j}C_{n-1}^{y_j-1}}{\theta_j^{y_j}(1-\theta_j)^{n-y_j}C_{n}^{y_j}}\notag\\
 	=&\theta_j\log{\frac{1}{\theta_j n}}+\frac{1}{n}\sum_{y_j=0}^{n}\theta_j^{y_j}(1-\theta_j)^{n-y_j}C_{n}^{y_j}y_j\log y_j\notag\\
 	\leq&\theta_j\log(\frac{1}{\theta_jn})+\theta_j\log(n\theta_j+(1-\theta_j))\label{applylemma}\\
 	=&\theta_j\log\left(1+\frac{1-\theta_j}{\theta_jn}\right)\notag\\
 	\leq&\frac{1-\theta_j}{n}\notag.
 	\end{align}
 	where (\ref{applylemma}) utilizes Lemma~\ref{prooflemma5}. For the second term, we can similarly obtain that:
 	\begin{equation*}
 	(1-\theta_j)\sum_{y_j=0}^{n-1}p(y_j\vert u_{ij}=0)\log\frac{p(y_j\vert u_{ij}=0)}{p(y_j)}\leq\frac{\theta_j}{n}
 	\end{equation*}
 	Therefore we have $I_{\theta}(\tilde{Y};X_i)\leq\frac{d}{n}$, and further $I_{\theta}(Y;U_i)\leq\frac{d}{n}$.
 \end{IEEEproof}

Proposition~\ref{mutual Gaussian} and \ref{mutual Bernoulli} show that the mutual information upper bounds of the minimax optimal estimation schemes scale with $O(\frac{1}{n})$ for the three models we consider, which means that we can achieve arbitrarily small mutual information as the number of users increases. Notice that we do not have to add noises or use random response as traditional digital schemes do to attain privacy. Our privacy guarantee comes from the additive nature of MAC channel. Since the processor only receives a noisy superposition of all signals, when the processor tries to infer a single user's original data, all the other users' signals become noises and provide privacy protection. Therefore, the over-the-air estimation schemes will satisfy the mutual information privacy constraints without sacrificing the utility of data. In other words, over-the-air estimation gives us free privacy due to its clever utilization of MAC channel, and that is a significant advantage over digital schemes.
\section{More Robust and General Private Estimation: Conditional Mutual Information Privacy}
Although the schemes we have proposed in our last section show excellent performance on mutual information bounds, the privacy guarantee is not so robust because it totally depends on other users' signals. If there are some users attacked and their data are leaked to the processor, the privacy protection for other users will also diminish. Therefore, we need to find a more robust privacy constraints to prevent such scenarios happen. Besides, the mutual information bounds for minimax optimal over-the-air schemes vary a lot for different families of distributions, and we want to find a more general estimation scheme applicable to all models. Next we will first propose a robust over-the-air estimation scheme which solves these two problems by simply adding local Gaussian noises and then we will prove that the new scheme still retains the advantages of lower estimation error compared to digital schemes. 
 
\subsection{Robust Over-the-Air Estimation Schemes}
To mitigate the negative impact of data leakage by other users, we utilize conditional mutual information $I_{\theta}(Y;U_i\vert U_1,\cdots,U_{i-1},U_{i+1},\cdots,U_n)$ as the privacy metric rather than mutual information. In this way, our privacy guarantee will no longer relies on other users' data. Notice that conditional mutual information constraints are stronger than mutual information privacy since $I_{\theta}(Y;U_i\vert U_1,\cdots,U_{i-1},U_{i+1},\cdots,U_n)\geq I_{\theta}(Y;U_i)$. To satisfy this new privacy constraint, we revise our primary minimax optimal over-the-air schemes by adding Gaussian noises locally to improve their robustness. We denote the minimax optimal estimation schemes (only consider deterministic and invertible encoding functions as we have done in last section) for $\mathcal{P}$ under power constraints $P$, channel noises variance $\sigma_0^2$ and the number of channel uses $s$ by $(\hat{\theta}^*(\mathcal{P},P,\sigma_0^2,s),f^*(\mathcal{P},P,\sigma_0^2,s))$, and the minimax optimal estimation error by $R^*(\mathcal{P},P,\sigma_0^2,s)$.Then we can formulate the robust over-the-air estimation schemes as follows:
\begin{definition}[$\epsilon$-Robust Over-the-Air Estimation Schemes]
Suppose the power constraints is $P$, channel noises variance is $\sigma_0^2$ and we have $s$ channel uses. Our estimation goal is the distribution family $\mathcal{P}$. Then the corresponding $\epsilon$-robust over-the-air estimation scheme is $(\hat{\theta}^*(\mathcal{P},P-\sigma_{pri}^2,\sigma_0^2+n\sigma_{pri}^2,s),f^*(\mathcal{P},P-\sigma_{pri}^2,\sigma_0^2+n\sigma_{pri}^2,s)+Z_{pri})$, where $\sigma_{pri}^2=\max\{\frac{sP-2\epsilon\sigma_0^2}{2\epsilon n+s},0\}$ and $Z_{pri}\sim\mathcal{N}(0,\sigma_{pri}^2I_s)$. 
\end{definition}

To characterize the property of $\epsilon$-robust over-the-air estimation schemes, we have the following theorem:
\begin{theorem}
\label{robust over}
Suppose the power constraints is $P$, channel noises variance is $\sigma_0^2$ with channel uses $s$ and the estimation goal is the distribution family $\mathcal{P}$. Then the $\epsilon$-robust over-the-air estimation scheme satisfies 
\begin{equation}
I_{\theta}(Y;U_i\vert U_1,\cdots,U_{i-1},U_{i+1},\cdots,U_n)\leq\epsilon,\forall 1\leq i\leq n,
\end{equation}
and the minimax estimation error is $R^*(\mathcal{P},P-\sigma_{pri}^2,\sigma_0^2+n\sigma_{pri}^2,s)$.
\end{theorem}
\begin{IEEEproof}
Let $\tilde{X}_i$ denote the output of $f^*(\mathcal{P},P-\sigma_{pri}^2,\sigma_0^2+n\sigma_{pri}^2,s)$ when the input is $U_i$. Then $X_i=\tilde{X}_i+Z_{pri,i}$, where $Z_{pri,i}\sim\mathcal{N}(0,\sigma_{pri}^2I_s)$. We can write the received signal $Y$ as follows:
\begin{equation*}
Y=\tilde{X_i}+\sum_{l\neq i}\tilde{X_l}+\sum_{i=1}^nZ_{pri,i}+Z=\tilde{X_i}+\sum_{l\neq i}\tilde{X_l}+Z_{total},
\end{equation*}
where $Z_{total}\sim\mathcal{N}(0,n\sigma_{pri}^2+\sigma_0^2)$. Denote $\tilde{X_i}+Z_{total}$ by $Y'$, then since  $f^*(\mathcal{P},P-\sigma_{pri}^2,\sigma_0^2+n\sigma_{pri}^2,s)$ is deterministic and invertible, we have 
\begin{align*}
&I_{\theta}(Y;U_i\vert U_1,\cdots,U_{i-1},U_{i+1},\cdots,U_n)\\
=&I_{\theta}(Y;\tilde{X_i}\vert \tilde{X}_1,\cdots,\tilde{X}_{i-1},\tilde{X}_{i+1},\cdots,\tilde{X}_n)\\
=&I_{\theta}(Y';\tilde{X_i})
\end{align*}

Since $Y'=\tilde{X_i}+Z_{total}$ and $\mathbb{E}[\tilde{X_i}]\leq P-\sigma_{pri}^2$, by Shannon's theorem we have:
\begin{equation*}
I_{\theta}(Y'_j;\tilde{X_{ij}})\leq\frac{1}{2}\log\left(\frac{P-\sigma_{pri}^2}{n\sigma_{pri}^2+\sigma_0^2}+1\right).
\end{equation*}

Therefore $I_{\theta}(Y;U_i\vert U_1,\cdots,U_{i-1},U_{i+1},\cdots,U_n)=\sum_jI_{\theta}(Y'_j;\tilde{X_{ij}})\leq\frac{s}{2}\log\left(\frac{P-\sigma_{pri}^2}{n\sigma_{pri}^2+\sigma_0^2}+1\right)$. Plugging in the expression of $\sigma_{pri}^2=\max\{\frac{sP-2\epsilon\sigma_0^2}{2\epsilon n+s},0\}$, we can obtain that $I_{\theta}(Y;U_i\vert U_1,\cdots,U_{i-1},U_{i+1},\cdots,U_n)\leq\epsilon$.

Meanwhile, notice that the estimation scheme is equivalent to estimate $\mathcal{P}$ under power constraint $P-\sigma_{pri}^2$ and channel noise variance $n\sigma_{pri}^2+\sigma_0^2$. Hence the corresponding estimation error is $R^*(\mathcal{P},P-\sigma_{pri}^2,\sigma_0^2+n\sigma_{pri}^2,s)$.
\end{IEEEproof}

Notice that during the proof of Theorem~\ref{robust over} we do not require the specific properties of distribution family $\mathcal{P}$. Thus our estimation scheme and its privacy guarantee hold for arbitrary distributions, which widens the application range of our scheme greatly.
 
\subsection{Minimax Estimation Error: Over-the-Air Vs Digital}
Now let us analyze the estimation performance of robust over-the-air estimation schemes. Combining Theorem~\ref{robust over} with Lemma~\ref{Gaussian minimax}, Lemma ~\ref{Bernoulli minimax} and Theorem~\ref{sparse minimax}, we can obtain the following corollaries which describe the estimation error of robust over-the-air schemes.
\begin{corollary}
For Gaussian location model, suppose the power constraints is $P$, channel noises variance is $\sigma_0^2$ and we have $s=d$ channel uses, then the minimax estimation error of $\epsilon$-robust over-the-air estimation scheme is:
\begin{equation}
	R_{robust}=\frac{d\sigma^2}{n}\left[1+\frac{d}{2n\epsilon}\left(1+\frac{B^2}{\sigma^2}\right)\right]
\end{equation}
\end{corollary}

\begin{corollary}
	For product Bernoulli model, suppose the power constraints is $P$, channel noises variance is $\sigma_0^2$ and we have $s=d$ channel uses.Then if $\sigma_0^2+n\sigma_{pri}^2\leq n^{\frac{3}{2}}(P-\sigma_{pri}^2)$, the minimax estimation error of $\epsilon$-robust over-the-air estimation scheme is:
	\begin{equation}
	R_{robust}=\frac{d}{4(\sqrt{n}+1)^2}(1+\frac{d}{2n\epsilon}),
	\end{equation}
	Otherwise,
	\begin{equation}
	R_{robust}=\frac{d}{4}\cdot\frac{1}{1+n\cdot\frac{2n\epsilon}{d}}.
	\end{equation}
\end{corollary}

\begin{corollary}
	For $m$-sparse Bernoulli model where $\frac{m}{d}\leq\frac{1}{2}$, suppose the power constraints is $P$, channel noises variance is $\sigma_0^2$ and we have $s=d$ channel uses.Then if $\frac{d^2}{4m(d-m)}(\sigma_0^2+n\sigma_{pri}^2)\leq n^{\frac{3}{2}}(P-\sigma_{pri}^2)$, the minimax estimation error of $\epsilon$-robust over-the-air estimation scheme is:
	\begin{align}
	&R_{robust}=\notag\\
	&\frac{m}{(\sqrt{n}+1)^2}\cdot(\frac{d-m}{d}
	+\frac{d}{2n\epsilon}\frac{d}{m(\sqrt{\frac{d-m}{m}}+\sqrt{\frac{m}{d-m}})^2}).
	\end{align}
	Otherwise,
	\begin{align}
&R_{robust}=\notag\\
&m\cdot\frac{1}{\frac{d}{d-m}+n\cdot\frac{2n\epsilon}{d}\frac{m(\sqrt{\frac{d-m}{m}}+\sqrt{\frac{m}{d-m}})^2}{d}}.
\end{align}
\end{corollary}

To display the advantages of our proposed schemes, we compare the scaling behavior of minimax estimation error under conditional mutual information constraint $I_{\theta}(Y;U_i\vert U_1,\cdots,U_{i-1},U_{i+1},\cdots,U_n)\leq\epsilon(\forall 1\leq i\leq n)$ of robust over-the-air schemes and order-optimal digital schemes \cite{barnes2020fisher} in Table~\ref{comparison}. We can see that our schemes reduce the estimation error generally by $O(\frac{1}{n})$ for the three models we consider, which is a significant performance enhancement. The reason why over-the-air estimation functions better is still the aggregation property of MAC channel. Although under conditional mutual information constraints we cannot use other users' data as interference, the locally added Gaussian noises of every user will aggregate at the receiver and the privacy-preserving effect is amplified by $n$ times, thus providing stronger privacy guarantee. Therefore, we can use much less noises locally than digital schemes and mitigate privacy's influence on data utility. 

\begin{table}
	\centering
	\caption{Comparison of the scaling behavior of estimation error attained by order-optimal digital schemes and robust over-the-air (ROA) schemes under conditional mutual information  constraints.} 
	\label{comparison}
	\setlength{\tabcolsep}{2mm}{
		\begin{tabular}{|c|c|c|c|}  
			\hline
			&digital schemes&ROA\\
			\hline
			Gaussian&$\frac{d^2\sigma^2}{n\epsilon}$&$\frac{d^2\sigma^2}{n^2\epsilon}$\\
			\hline
			Bernoulli&$\frac{d^2}{n\epsilon}$&$\frac{d^2}{n^2\epsilon^2}$\\
			\hline
			$m$-Sparse Bernoulli&$\frac{m^2\log d}{n\epsilon}$&$\frac{md}{n^2\epsilon}$\\
			($\frac{m}{d}\leq\frac{1}{2}$)&$(n\epsilon\geq d\log d)$&\\
			\hline
		\end{tabular}
	}
\end{table}

\section{Conclusion}
In this paper we study the distributed statistical estimation problem under privacy constraints. We first derive the minimax optimal over-the-air estimation schemes for three common models (Gaussian location model, product Bernoulli model and $m$-sparse Bernoulli model) and show that our schemes can obtain free mutual information privacy without sacrificing the data utility. This is because over-the-air transmission can utilize the additive nature of MAC channel and other users' signals become noises when the processor tries to infer a single user's data. Then we consider conditional mutual information privacy and propose robust over-the-air estimation schemes which can obtain more robust privacy guarantee and apply to more general models. The comparison of estimation error between our proposed schemes and traditional digital transmission schemes further displays the superiority of over-the-air estimation in such setting.

\bibliography{Reference}

\begin{thebibliography}{10}
\providecommand{\url}[1]{#1}
\csname url@samestyle\endcsname
\providecommand{\newblock}{\relax}
\providecommand{\bibinfo}[2]{#2}
\providecommand{\BIBentrySTDinterwordspacing}{\spaceskip=0pt\relax}
\providecommand{\BIBentryALTinterwordstretchfactor}{4}
\providecommand{\BIBentryALTinterwordspacing}{\spaceskip=\fontdimen2\font plus
\BIBentryALTinterwordstretchfactor\fontdimen3\font minus
  \fontdimen4\font\relax}
\providecommand{\BIBforeignlanguage}[2]{{%
\expandafter\ifx\csname l@#1\endcsname\relax
\typeout{** WARNING: IEEEtran.bst: No hyphenation pattern has been}%
\typeout{** loaded for the language `#1'. Using the pattern for}%
\typeout{** the default language instead.}%
\else
\language=\csname l@#1\endcsname
\fi
#2}}
\providecommand{\BIBdecl}{\relax}
\BIBdecl

\bibitem{xu2019federated}
J.~Xu and F.~Wang, ``Federated learning for healthcare informatics,''
  \emph{arXiv preprint arXiv:1911.06270}, 2019.

\bibitem{hegedHus2019decentralized}
I.~Heged{\H{u}}s, G.~Danner, and M.~Jelasity, ``Decentralized recommendation
  based on matrix factorization: A comparison of gossip and federated
  learning,'' in \emph{Joint European Conference on Machine Learning and
  Knowledge Discovery in Databases}.\hskip 1em plus 0.5em minus 0.4em\relax
  Springer, 2019, pp. 317--332.

\bibitem{hard2018federated}
A.~Hard, K.~Rao, R.~Mathews, S.~Ramaswamy, F.~Beaufays, S.~Augenstein,
  H.~Eichner, C.~Kiddon, and D.~Ramage, ``Federated learning for mobile
  keyboard prediction,'' \emph{arXiv preprint arXiv:1811.03604}, 2018.

\bibitem{duchi2013local}
J.~C. Duchi, M.~I. Jordan, and M.~J. Wainwright, ``Local privacy and
  statistical minimax rates,'' in \emph{2013 IEEE 54th Annual Symposium on
  Foundations of Computer Science}.\hskip 1em plus 0.5em minus 0.4em\relax
  IEEE, 2013, pp. 429--438.

\bibitem{sheffet2018locally}
O.~Sheffet, ``Locally private hypothesis testing,'' \emph{arXiv preprint
  arXiv:1802.03441}, 2018.

\bibitem{acharya2018distributed}
J.~Acharya, C.~L. Canonne, and H.~Tyagi, ``Distributed simulation and
  distributed inference,'' \emph{arXiv preprint arXiv:1804.06952}, 2018.

\bibitem{barnes2020fisher}
L.~P. Barnes, W.-N. Chen, and A.~Ozgur, ``Fisher information under local
  differential privacy,'' \emph{arXiv preprint arXiv:2005.10783}, 2020.

\bibitem{erlingsson2014rappor}
{\'U}.~Erlingsson, V.~Pihur, and A.~Korolova, ``Rappor: Randomized aggregatable
  privacy-preserving ordinal response,'' in \emph{Proceedings of the 2014 ACM
  SIGSAC conference on computer and communications security}, 2014, pp.
  1054--1067.

\bibitem{kairouz2016discrete}
P.~Kairouz, K.~Bonawitz, and D.~Ramage, ``Discrete distribution estimation
  under local privacy,'' \emph{arXiv preprint arXiv:1602.07387}, 2016.

\bibitem{ye2018optimal}
M.~Ye and A.~Barg, ``Optimal schemes for discrete distribution estimation under
  locally differential privacy,'' \emph{IEEE Transactions on Information
  Theory}, vol.~64, no.~8, pp. 5662--5676, 2018.

\bibitem{cormode2018privacy}
G.~Cormode, S.~Jha, T.~Kulkarni, N.~Li, D.~Srivastava, and T.~Wang, ``Privacy
  at scale: Local differential privacy in practice,'' in \emph{Proceedings of
  the 2018 International Conference on Management of Data}, 2018, pp.
  1655--1658.

\bibitem{makhdoumi2014information}
A.~Makhdoumi, S.~Salamatian, N.~Fawaz, and M.~M{\'e}dard, ``From the
  information bottleneck to the privacy funnel,'' in \emph{2014 IEEE
  Information Theory Workshop (ITW 2014)}.\hskip 1em plus 0.5em minus
  0.4em\relax IEEE, 2014, pp. 501--505.

\bibitem{calmon2015fundamental}
F.~P. Calmon, A.~Makhdoumi, and M.~M{\'e}dard, ``Fundamental limits of perfect
  privacy,'' in \emph{2015 IEEE International Symposium on Information Theory
  (ISIT)}.\hskip 1em plus 0.5em minus 0.4em\relax IEEE, 2015, pp. 1796--1800.

\bibitem{asoodeh2015maximal}
S.~Asoodeh, F.~Alajaji, and T.~Linder, ``On maximal correlation, mutual
  information and data privacy,'' in \emph{2015 IEEE 14th Canadian workshop on
  information theory (CWIT)}.\hskip 1em plus 0.5em minus 0.4em\relax IEEE,
  2015, pp. 27--31.

\bibitem{liao2017hypothesis}
J.~Liao, L.~Sankar, V.~Y. Tan, and F.~du~Pin~Calmon, ``Hypothesis testing under
  mutual information privacy constraints in the high privacy regime,''
  \emph{IEEE Transactions on Information Forensics and Security}, vol.~13,
  no.~4, pp. 1058--1071, 2017.

\bibitem{sreekumar2019optimal}
S.~Sreekumar and D.~G{\"u}nd{\"u}z, ``Optimal privacy-utility trade-off under a
  rate constraint,'' in \emph{2019 IEEE International Symposium on Information
  Theory (ISIT)}.\hskip 1em plus 0.5em minus 0.4em\relax IEEE, 2019, pp.
  2159--2163.

\bibitem{wang2016relation}
W.~Wang, L.~Ying, and J.~Zhang, ``On the relation between identifiability,
  differential privacy, and mutual-information privacy,'' \emph{IEEE
  Transactions on Information Theory}, vol.~62, no.~9, pp. 5018--5029, 2016.

\bibitem{cuff2016differential}
P.~Cuff and L.~Yu, ``Differential privacy as a mutual information constraint,''
  in \emph{Proceedings of the 2016 ACM SIGSAC Conference on Computer and
  Communications Security}, 2016, pp. 43--54.

\bibitem{amiri2020machine}
M.~M. Amiri and D.~G{\"u}nd{\"u}z, ``Machine learning at the wireless edge:
  Distributed stochastic gradient descent over-the-air,'' \emph{IEEE
  Transactions on Signal Processing}, vol.~68, pp. 2155--2169, 2020.

\bibitem{seif2020wireless}
M.~Seif, R.~Tandon, and M.~Li, ``Wireless federated learning with local
  differential privacy,'' \emph{arXiv preprint arXiv:2002.05151}, 2020.

\end{thebibliography}
\begin{appendices}
\section{Proof of Lemma~\ref{prooflemma1}}
First consider the case where $\alpha\geq\frac{1}{2(n-\sqrt{n})C}$. If $\beta\geq\frac{1}{2}$, then $\sup_{\theta}R(\theta;f_C,\hat{\theta}_{\alpha,\beta})=[(2nC\alpha-1)(2\beta-1)]m+d[\alpha^2\sigma_0^2+(\beta-nC\alpha)^2]$. By some calculations, we know the best choice is as follows:
\begin{equation*}
\alpha_{1,1}=\frac{1}{2C(n-\sqrt{n})},
\end{equation*}
\begin{equation*}
\beta_{1,1}=
\begin{cases}
\frac{1}{2},& if \frac{m}{d}\geq\frac{1}{2},\\
(1-\frac{2m}{d})nC\alpha+\frac{m}{d},& if \frac{m}{d}\leq\frac{1}{2}
\end{cases}
\end{equation*}
If $\beta\leq\frac{1}{2}$, then $\sup_{\theta}R(\theta;f_C,\hat{\theta}_{\alpha,\beta})=d[\alpha^2\sigma_0^2+(\beta-nC\alpha)^2]$. And the best choice is:
\begin{equation*}
\alpha_{1,2}=\frac{1}{2C(n-\sqrt{n})},
\end{equation*}
\begin{equation*}
\beta_{1,2}=\frac{1}{2}.
\end{equation*}
We can see that the optimal scheme for the second case is included in the first case. Thus the optimal scheme here should be:
\begin{equation*}
\alpha_{1}=\alpha_{1,1},
\end{equation*}
\begin{equation*}
\beta_{1}=\beta_{1,1}.
\end{equation*}

Then consider the case where $\alpha\leq\frac{1}{2(n+\sqrt{n})C}$. If $\beta\leq\frac{1}{2}$, then $\sup_{\theta}R(\theta;f_C,\hat{\theta}_{\alpha,\beta})=[(2nC\alpha-1)(2\beta-1)]m+d[\alpha^2\sigma_0^2+(\beta-nC\alpha)^2]$. By some calculations, we know the best choice is as follows:

when $\frac{m}{d}\leq\frac{1}{2}$,
\begin{equation*}
\alpha_{2,1}=\min\{\frac{1}{2C(n+\sqrt{n})},\frac{2m(d-m)nC}{d^2\sigma_0^2+4m(d-m)n^2C^2)}\},
\end{equation*}
\begin{equation*}
\beta_{2,1}=(1-\frac{2m}{d})nC\alpha+\frac{m}{d}.
\end{equation*}

when $\frac{m}{d}\geq\frac{1}{2}$,
\begin{equation*}
\alpha_{2,1}=\min\{\frac{1}{2C(n+\sqrt{n})},\frac{nC}{2(\sigma_0^2+n^2C^2)}\},
\end{equation*}
\begin{equation*}
\beta_{2,1}=\frac{1}{2}.
\end{equation*}

If $\beta\geq\frac{1}{2}$, then $\sup_{\theta}R(\theta;f_C,\hat{\theta}_{\alpha,\beta})=d[\alpha^2\sigma_0^2+(\beta-nC\alpha)^2]$. And the best choice is:
\begin{equation*}
\alpha_{2,2}=\min\{\frac{1}{2C(n+\sqrt{n})},\frac{nC}{2(\sigma_0^2+n^2C^2)}\},
\end{equation*}
\begin{equation*}
\beta_{2,2}=\frac{1}{2}.
\end{equation*}
Again, the second case is included in the first one. Thus the optimal scheme here should be:
\begin{equation*}
\alpha_{2}=\alpha_{2,1},
\end{equation*}
\begin{equation*}
\beta_{2}=\beta_{2,1}
\end{equation*}

Now let's compare the performance of $\alpha_1,\beta_1$ and $\alpha_2,\beta_2$. Notice that $\frac{2m(d-m)nC}{d^2\sigma_0^2+4m(d-m)n^2C^2)}<\frac{1}{2(n-\sqrt{n})C}+\frac{1}{2(n+\sqrt{n})C}$, after some calculations, we have:
\begin{equation*}
\sup_{\theta}R(\theta;f_C,\hat{\theta}_{\alpha_1,\beta_1})>\sup_{\theta}R(\theta;f_C,\hat{\theta}_{\alpha_2,\beta_2}).
\end{equation*}

Thus the optimal scheme is $\hat{\theta}_{\alpha_2,\beta_2}$, as stated in the lemma.
\section{Proof of Lemma~\ref{prooflemma2}}
By variable substitution, the minimax risk is equivalent to the expression below:
\begin{align*}
&\frac{1}{d}\sup_{\theta}R(\theta;f_C,\hat{\theta}_{\alpha,\beta})\\
=&\sup_{\theta'}L(\theta';f_C,\hat{\theta}_{\alpha,\beta})\\
=&[4(n^2-n)C^2\alpha^2-4nC\alpha+1]\theta'^2\notag\\
&+[4nC^2\alpha^2+4nC\alpha\beta-2\beta-4n^2C^2\alpha^2+2nC\alpha]\theta'\notag\\
&+[\alpha^2\sigma^2+(\beta-nC\alpha)^2].
\end{align*}
where $0\leq\theta'\leq\frac{m}{d}$.

When $\frac{m}{d}\geq\frac{1}{2}$, we have:
\begin{align*} 
&L(\frac{1}{2};f_C,\alpha,\beta)\\
=&\alpha^2(nC^2+\sigma_0^2)+(\beta-\frac{1}{2})^2\\
\geq&\alpha^{*2}(nC^2+\sigma_0^2)\\
=&\sup_{\theta'}L(\theta';f_C,\alpha^*,\beta^*)
\end{align*}

When $\frac{m}{d}\leq\frac{1}{2}$, we have:
\begin{align*} 
&L(\frac{m}{d};f_C,\alpha,\beta)\\
\geq&L(\frac{m}{d};f_C,\alpha,(1-\frac{2m}{d})nC\alpha+\frac{m}{d})\\
\geq&\alpha^2[4nC^2\frac{m}{d}(1-\frac{m}{d})+\sigma_0^2]\\
\geq&\alpha^{*2}[4nC^2\frac{m}{d}(1-\frac{m}{d})+\sigma_0^2]\\
\end{align*}

Notice that $\frac{2nC^2\alpha^2-\frac{m}{d}(2nC\alpha-1)^2}{4nC^2\alpha^2-(2nC\alpha-1)^2}\geq\frac{m}{d}$. Since $4(n^2-n)C^2\alpha^2-4nC\alpha+1\leq0$,we have $\sup_{\theta'}L(\theta';f_C,\alpha^*,\beta^*)=L(\frac{m}{d};f_C,\alpha^*,\beta^*)$. Therefore in this case we also have $L(\frac{m}{d};f_C,\alpha,\beta)\geq\sup_{\theta'}L(\theta';f_C,\alpha^*,\beta^*)$.
\section{Proof of Lemma~\ref{prooflemma4}}
Without loss of generality assume that $B>A$. From Lemma~\ref{prooflemma3}, we know the equivalent $C$ for $f$ is $C=\frac{B-A}{2}$, then $A=B-2C$. Thus for power constraints we have:
\begin{equation*}
\sum_{i=1}^d\theta_i(A+2C)^2+\sum_{i=1}^d(1-\theta_i)A^2\leq dP
\end{equation*}
which is equivalent to
\begin{equation*}
(4AC+4C^2)\sum_{i}\theta_i+dA^2-dP\leq 0, \forall 0\leq\sum_i\theta_i\leq m
\end{equation*}
This is a linear function of $\sum_i\theta_i$, thus is equivalent to:
\begin{equation}
\label{power inequality1}
A^2\leq P,
\end{equation}
\begin{equation}
\label{power inequality2}
4mC^2+4mAC+d(A^2-P)\leq 0.
\end{equation}
Since we want to maximize $C$, from inequality~(\ref{power inequality2}) we know that:
\begin{equation}
\label{max c}
C=-\frac{A}{2}+\sqrt{\frac{A^2}{4}+\frac{d}{4m}(P-A^2)}
\end{equation}	
Considering the constraint (\ref{power inequality1}), we can maximize (\ref{max c}) and obtain the results in the lemma.
\section{Proof of Lemma~\ref{prooflemma5}}
For any $t>0$, we have $\ln\frac{v}{t}\leq\frac{v}{t}-1$ for all $v>0$, thus $v\ln v\leq\frac{v^2}{t}+v\ln\frac{t}{e}$.
Choose $t=\frac{\omega^2+\mu^2}{\mu}$,then we can obtain that $\mathbb{E}[V\ln V]\leq\mu\ln\frac{\omega^2+\mu^2}{\mu}$.
\end{appendices}
\end{document}